\documentclass{aa}  

\usepackage{graphicx}
\usepackage{txfonts}
\usepackage{color}
\usepackage{multirow}

\usepackage{hyperref}

\makeatletter
\renewcommand*\aa@pageof{, page \thepage{} of \pageref*{LastPage}}
\makeatother

\begin{document} 

\title{Cloud structure and young star distribution in the Dragonfish complex}

\author{Nestor Sanchez$^{1}$
        \and
        Elisa Nespoli$^{1}$
        \and
        Marta Gonzalez$^{1}$
        \and
        Juan B. Climent$^{1,2}$}

\titlerunning{Cloud structure and young star distribution in the Dragonfish complex}

\authorrunning{Sanchez et al.}

\institute{$^{1}$Universidad Internacional de Valencia (VIU),
           C/Pintor Sorolla 21, E-46002 Valencia, Spain\\
           $^{2}$Departament d'Astronomia i Astrof\'{i}sica, Universitat de Val\`{e}ncia, Burjassot, E-46100, Spain}

\date{Received ...; accepted ...}

\abstract
{Star formation is a complex process involving several physical mechanisms interacting with each other at different spatial scales. One way to shed some light on this process is to analyse the relationship between the spatial distributions of gas and newly-formed stars. In order to obtain robust results, it is necessary for this comparison to be made using quantitative and consistent descriptors applied to the same star-forming region.}
{Here, we use fractal analysis to characterise and compare in a self-consistent way the structure of the cloud and the distribution of young stellar objects (YSO) in the Dragonfish star-forming complex.}
{Different emission maps of the Dragonfish Nebula were retrieved from the NASA/IPAC Infrared Science and the Planck Legacy archives. Moreover, we used photometric information from the AllWISE catalogue to select a total of 1082 YSOs in the region, for some of which we derived physical properties from their spectral energy distributions (SEDs). For both datasets (cloud images and YSOs), the three-dimensional fractal dimension ($D_f$) was calculated using previously developed and calibrated algorithms.}
{The fractal dimension of the Dragonfish Nebula ($D_f = 2.6-2.7$) agrees very well with values previously obtained for the Orion, Ophiuchus, and Perseus clouds. On the other hand, YSOs exhibit on average a significantly smaller value ($D_f = 1.9-2.0$) that indicates a much more clumpy structure than the material from which they formed. Younger Class I and Class II sources have smaller values ($D_f = 1.7 \pm 0.1$) than more evolved Transition Disk objects ($D_f = 2.2 \pm 0.1$), evidencing a certain evolutionary effect where an initially clumpy structure tends to gradually disappear over time.}
{The Dragonfish complex exhibits a structure similar to that of other molecular clouds in the Galaxy. However, we have found clear and direct  evidence that the clustering degree of the newly born stars is significantly higher than that of the parent cloud from which they formed. The physical mechanism behind this behaviour is still not clear.}

\keywords{ISM: clouds --
          ISM: structure --
          ISM: individual objects: Dragonfish Nebula --
          Stars: early-type --
          Stars: formation}

\maketitle

\section{Introduction}

Star formation is a complex process that is still not fully understood. There is an accepted general picture in which gas and dust inside giant molecular clouds (GMCs) gravitationally collapse to form groups of protostars. After this, protostellar winds and jets blow away the surrounding clouds leaving behind clusters of newly formed stars \citep[see, for instance, the review by][]{McKee2007}. However, the details of the process are much more complex than this. The internal structure of GMCs is mainly driven by turbulent motions whose origin is still under debate \citep{Larson1981,Elmegreen2004,Smith2022}. Turbulence tends to act against the gravitational collapse, but it can also originate shocks and high-density regions promoting the collapse. Apart from turbulence and self-gravitation, there are other physical mechanisms such as magnetic fields, thermal pressure and radiation fields that may play important roles at different spatial scales and in different moments of the star formation process \citep{McKee2007}. Stellar feedback from newly formed stars injects energy into the medium that can either disperse the gas (preventing the formation of other stars) or compress it (triggering the formation of additional stars) depending on many different factors \citep{Bally2016}. Even if very few physical mechanisms were considered, the interaction among all the involved processes and the interaction among different regions of the GMC at different spatial scales convert star formation in a highly non-linear chaotic process, in the sense of being very sensitive to small variations on the initial and environmental conditions \citep{Sanchez1999,Jaffa2022}.

The study of the star formation process may be divided into two steps. Firstly, one needs to address the initial distribution of gas and dust in GMCs, i.e. the initial conditions of the process. Secondly, one can focus on the way and degree in which this initial distribution is transferred or converted into new-born stars, i.e. the star formation process itself. Each one of these parts is a complex research line with many physical processes and many observational problems involved. A way to yield some light into the problem is using a two-sided approach. On the one hand, to study and characterise in detail the structure and properties of GMCs that represent the initial conditions of the star formation process. On the other hand, to analyse the distribution and properties of YSOs. The comparison of these two parts may help to understand the process from which the parental cloud is transformed into newly-formed stars. In order to draw solid conclusions, this comparison should be made for the same star-forming region and using quantitative and consistent descriptors.  In the literature, several descriptors and techniques have been considered for the characterisation of the internal structure of interstellar clouds, such as structure tree methods \citep{Houlahan1992}, Delta-variance techniques \citep{Stutzki1998,Elia2014,Dib2020}, principal component analysis \citep{Ghazzali1999}, metric space techniques \citep{Khalil2002,Robitaille2010}, dendrograms \citep{Rosolowsky2008,Colombo2015}, convolutional neural networks \citep{Bates2020,Bates2023}, and fractal \citep{Falgarone1991,Vogelaar1994,Sanchez2005,Sanchez2007a,Lee2016,Marchuk2021} and multifractal \citep{Chappell2001,Khalil2006,Elia2018} analysis, among others. Regarding the structure of the distribution of formed stars and star clusters, commonly used methods include simple kernel density estimators \citep{Silverman1986}, the nearest neighbour distribution and the two-point correlation function \citep[see, for instance,][]{Gomez1993,Larson1995,Simon1997,Hartmann2002,Kraus2008}. The correlation function may be used to directly estimate the fractal correlation dimension of star and cluster distributions \citep{deLaFuenteMarcos2006,Kraus2008,Sanchez2007b,Sanchez2009}. A different method was introduced by \citet{Cartwright2004} which proposed the use of the so-called $Q$-parameter, calculated from the minimum spanning tree, to quantify the spatial substructure. This method has the advantage of being able to distinguish between centrally concentrated and fractal-like distributions, and it has been widely used in different star-forming regions \citep[see, for example,][and references therein]{Jaffa2017,Hetem2019}. Other more recent techniques or variants of already established methods to characterise the internal structure of interstellar clouds and young stars or star clusters include the INDICATE tool \citep{Buckner2019,BaylockSquibbs2022}, the S2D2 procedure \citep{Gonzalez2021}, the Moran's I statistic \citep{Arnold2022}, and the RJ Plots \citep{Jaffa2018,Clarke2022}.

Each method has its advantages, disadvantages, and limitations, and the choice of which to use depends, among other factors, on the scientific goal to be addressed. Fractal analysis is a particularly suitable tool because of the observed hierarchical and self-similar structure of the interstellar medium, which resembles a fractal system \citep[see][and references therein]{Bergin2007}. In fractal analysis, the degree of spatial heterogeneity can be quantified through a simple parameter: the fractal dimension ($D_f$). One advantage of this approach is that $D_f$ can be calculated for both continuous and discrete structures, which allows a direct comparison between the distribution of gas and dust in the parental cloud and the distribution of newly-formed stars. It is believed that very young stars and clusters should follow the fractal patterns of the interstellar medium from which they formed but that such patterns could be dissipated on short timescales \citep{Goodwin2004,Sanchez2009,Allison2010,Sun2022}. However, it is not clear whether the wide variety of observed spatial patterns is due to differences in the structure of the original clouds or to evolutionary or environmental effects \citep{Goodwin2004,Allison2010,Ballone2020,Daffern2020,Gonzalez2021}.

In this work, we use fractal analysis to examine in a systematic and consistent way the distribution of gas and YSOs in the Dragonfish region. Dragonfish (G298.4-0.4) is a star-forming complex located at $(l,b) = (298,0.4)$~deg first detected by \citet{Russeil1997} as a clump of HII regions at 10 kpc. Some authors suggested that the Dragonfish complex contained a supermassive OB association \citep{Rahman2011a,Rahman2011b}. However, a detailed work by \citet{delaFuente2016} showed that such an association does not exist and that the existing young massive clusters and Wolf-Rayet stars can explain most of the observed ionisation. \citet{delaFuente2016} estimated that this region is located at the outer edge of the Sagittarius-Carina spiral arm, at a distance of $d = 12.4$~kpc. However, \citet{Rate2020} found a much closer distance of 5.2~kpc using data from the Gaia DR2, although they warned that their distance estimation may be inaccurate. Previous studies suggest that Dragonfish is among the largest and most massive cloud complexes in the Milky Way \citep[see][and refernces therein]{delaFuente2016}, which makes it an interesting region to investigate the star formation process.
Section~\ref{sec_gas} of this work is dedicated to characterise the distribution of gas and dust in the Dragonfish nebula. In Section~\ref{sec_yso} we use photometric information in several bands to search for YSOs, 
determine their physical properties, and study the spatial and hierarchical clustering of the selected YSOs.
A comparison between the distributions of gas and YSOs is given in Section~\ref{sec_discusion} and, finally, Section~\ref{sec_conclusion} summarises our main conclusions.

\section{Gas and dust distribution}
\label{sec_gas}

\subsection{Used data}

\begin{figure*}
\centering
\includegraphics[width=\textwidth]{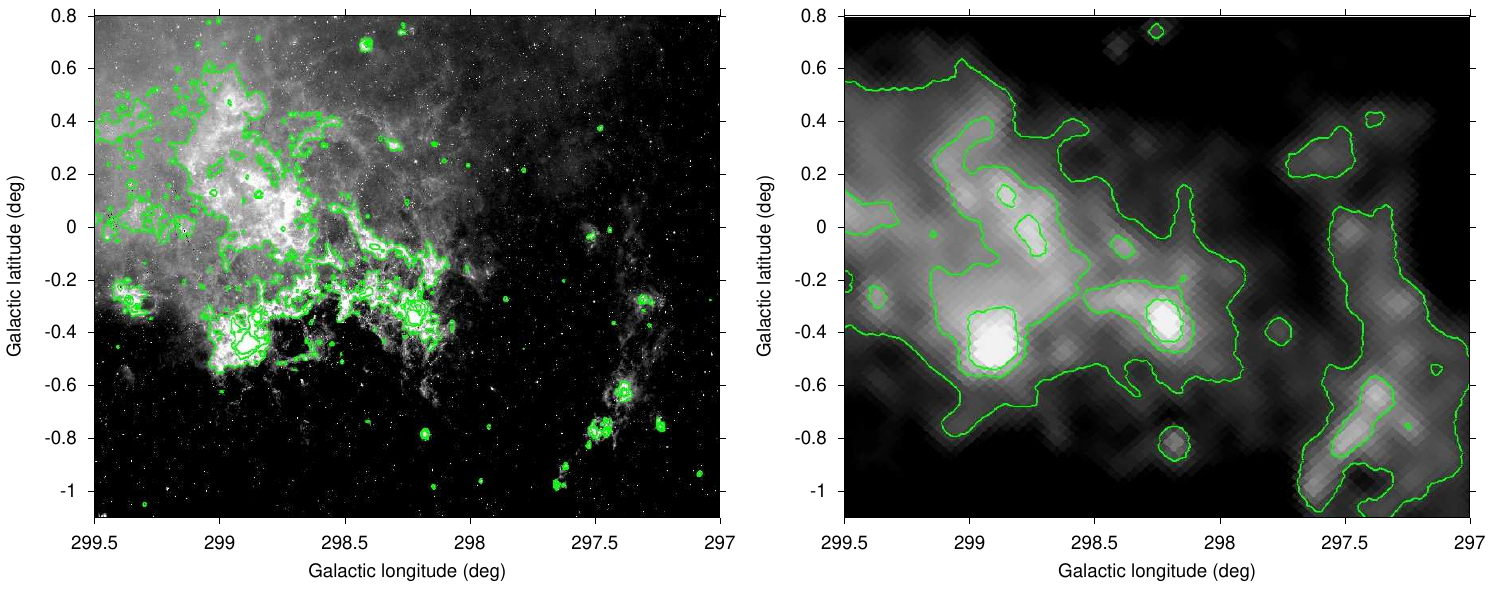}
\caption{Images of the Dragonfish Nebula covering the region studied in this work. \textit{Left panel}: Mosaic obtained by assembling 27 Spitzer IRAC images in the 8~$\mu$m band. \textit{Right panel}: map retrieved from the Planck Legacy Archive (HFI 545 GHz data). Both maps are in logarithmic grey scale and, for reference, three contours have been drawn at 30, 60 and 100 MJy/sr (Spitzer map) and at 100, 150 and 200 MJy/sr (Planck map).}
\label{fig_mapas}
\end{figure*}

We selected a region large enough to cover the entire Dragonfish star-forming complex. The region is defined by the galactic coordinates $l = (297.0, 299.5)$ deg and $b = (-1.1, +0.8)$ deg. By using the NASA/IPAC Infrared Science Archive\footnote{\url{https://irsa.ipac.caltech.edu}}, we downloaded images from the Infrared Array Camera (IRAC) of the Spitzer mission \citep{spitzer2004} and created a mosaic of the region using the Montage program\footnote{\url{http://montage.ipac.caltech.edu/}}. We performed this process for the four IRAC channels. The obtained results did not show significant differences among the four channels, so we present here the results using the $8~\mu$m IRAC channel only, namely channel 4. We also searched for additional data in other wavelengths but radio maps usually do not have enough spatial resolution to achieve our science goals. A relatively good enough image at microwave wavelengths was obtained from the Planck mission \citep{planck2011}. In particular, we downloaded from the Planck Legacy Archive\footnote{\url{https://pla.esac.esa.int}} the map of the HFI 545~GHz channel ($550~\mu$m). Figure~\ref{fig_mapas} displays the maps from Spitzer and Planck used in this study. 
In general, these maps trace the distribution of gas and dust in the Dragonfish Nebula. The Planck map is basically a thermal dust emission map \citep{Planck2016} whereas IRAC channel 4 is dominated by polycyclic aromatic hydrocarbon (PAH) emission \citep{Draine2007}, which tends to spatially correlate with molecular gas \citep{Schinnerer2013}.

\subsection{Fractal dimension}

In this section, we use fractal analysis to study the distribution of gas and dust in the Dragonfish region. This tool uses only one parameter, the fractal dimension $D_f$, to characterise the manner in which the gas is distributed. A $D_f$ value of 3 indicates a homogeneous three-dimensional spatial distribution, while progressively smaller values of $D_f$ correspond to increasingly irregular distributions with higher degrees of clumpiness \citep{Mandelbrot1983}. Monofractal clouds can be characterised by a single $D_f$ value that is valid across the entire range of spatial scales over which the gas is distributed. Although some evidence of multifractality in the interstellar medium (ISM) has been reported \citep{Chappell2001}, 
this remains an open issue, and a systematic analysis assuming a nearly monofractal behaviour may still provide valuable insights into the underlying structure of the ISM.

In general, interstellar clouds are observed as two-dimensional images projected onto the celestial sphere. Therefore, to study the fractal properties of clouds, many authors use the so-called perimeter-area relation to calculate the dimension of the contours of the projected clouds
\citep{Bazell1988,Dickman1990,Falgarone1991,Hetem1993,Vogelaar1994,Sanchez2005,Marchuk2021}, that we denote as $D_{per}$. In general, the relation between 2D and 3D fractal dimension values is not trivial but in principle $D_{per}$ can only vary between the theoretical limits of $D_{per}=1$ for the case of smooth projected contours and $D_{per}=2$ for extremely irregular contours \citep[see discussions in][]{Beattie2019,Sanchez2005}. In previous works \citep{Sanchez2005,Sanchez2007a}, we implemented and optimised an algorithm to estimate $D_{per}$ in a reliable way in cloud emission maps.
The method defines objects as sets of connected pixels with intensity values above a defined threshold. In order to increase the number of objects, the algorithm uses $\sim$20 brightness levels equally spaced between the minimum and the maximum brightness of the map. Several tests performed with both simulations and real maps showed that the obtained $D_{per}$ values do not depend on the exact number of brightness levels as long as they are not too few ($\lesssim 5$). 
Then, the perimeter and area of each object in the image are calculated and the best linear fit is determined in a log(perimeter)-log(area) plot, being $D_{per}/2$ the slope of the fit \citep{Mandelbrot1983}. The algorithm was optimised to account for problems occurring at the image edges as well as signal-to-noise ratio and resolution effects. More importantly, this algorithm was used to characterise in detail the relationship between $D_{per}$ and $D_f$.
The relation $D_{per} - D_f$ was empirically determined by simulating three-dimensional clouds with well-defined fractal dimensions and projecting them onto random planes. In general and as expected, $D_{per}$ decreases (more convoluted boundaries) as $D_f$ increases (more irregular and fragmented clouds). However, the exact relation is not a simple function and, as the image resolution decreases, there is a tendency of $D_{per}$ to decrease because the details of the roughness disappear as the pixel size increases. The calculated functional forms relating $D_{per}$, $D_f$ and $N_{pix}$ (the maximum object size in pixel units) were presented in Fig.~8 and Table~1 of \citet{Sanchez2005} and are used here to estimate $D_f$ for the Dragonfish Nebula. 

\begin{figure}
\centering
\includegraphics[width=0.5\textwidth]{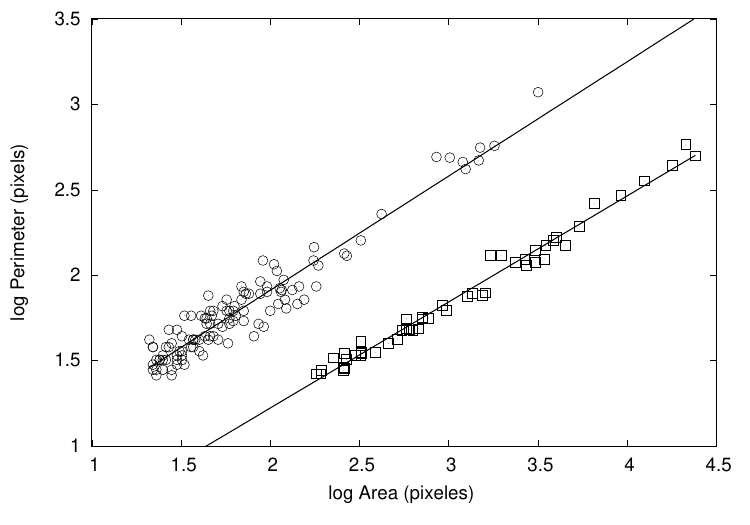}
\caption{Perimeter as a function of the area in pixels for objects (clouds) in the Dragonfish Nebula using the Spitzer (open circles) and the Planck (open square) maps. Planck data have been vertically shifted by $-0.5$ for clarity. Solid lines are the corresponding best linear fits for all the data points.}
\label{fig_Dper}
\end{figure}

We applied the previously described algorithm to the maps shown in Fig.~\ref{fig_mapas}. The obtained perimeter-area relations are shown in Fig.~\ref{fig_Dper}.
The corresponding fractal dimension values are summarised in Table~\ref{tab_dper}, where $D_f$ is estimated from $D_{per}$ based on the simulations of projected clouds performed in \citet{Sanchez2005}.
\begin{table}
\caption{Perimeter-area based dimension ($D_{per}$) and three-dimensional fractal dimension ($D_f$) for the Spitzer and Planck maps shown in Fig.~\ref{fig_mapas}.}
\label{tab_dper}
\centering
\begin{tabular}{c c c c}
\hline
Map & Pixels & $D_{per}$ & $D_f$ \\
\hline
Spitzer & $1525 \times 939$ & $1.34 \pm 0.04$ & $2.6-2.7$ \\
Planck & $697 \times 549$ &  $1.24 \pm 0.02$ & $2.6-2.7$ \\
\hline
\end{tabular}
\end{table}
The Spitzer map we are using has a pixel size corresponding to the ``good resolution'' case in \citet[][see Fig.~8 and Table~1 in this paper]{Sanchez2005}, i.e. the case with $N_{pix} \geq 400$ where $N_{pix}$ is the maximum object size in pixel units. In contrast, the Planck map has a pixel size corresponding to the case $N_{pix} \simeq 200$. Thus, the relatively small $D_{per}$ value of the Planck map in Table \ref{tab_dper} is due to resolution effects that tend to smooth the contours. For this reason, after correcting for resolution effects, the three-dimensional fractal dimensions result it the same value for both maps. The obtained value $D_f = 2.6-2.7$ for the Dragonfish Nebula agrees with our previous results for emission maps in different molecular lines of the Orion, Ophiuchus, and Perseus clouds, where the fractal dimensions are always in the range $2.6 \lesssim D_f \lesssim 2.8$ \citep{Sanchez2005,Sanchez2007a}.
These $D_f$ values are significantly higher than the average value $D_f \simeq 2.3$ commonly assumed for the ISM \citep{Elmegreen2001}.

\section{Young stellar object candidates}
\label{sec_yso}

\subsection{Candidate selection}

\begin{figure}
\centering
\includegraphics[width=0.5\textwidth]{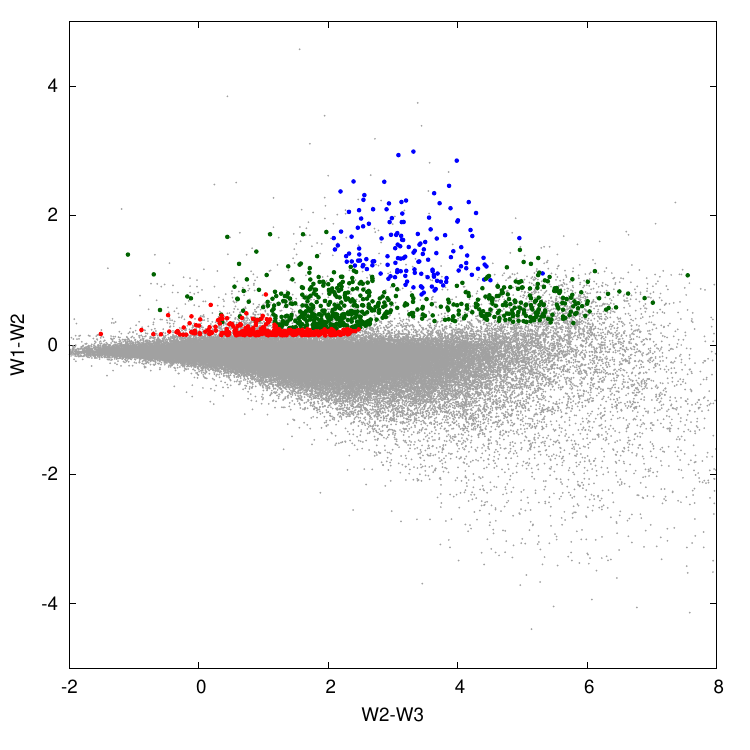}
\caption{Colour-colour diagram, $W1-W2$ versus $W2-W3$, for all stars in the sample (grey dots) and for sources fulfilling criteria of Class I (blue dots), Class II (green dots) and Transition disk (red dots) objects.}
\label{fig_yso_colorcolor}
\end{figure}

Within the selected region, we used VizieR\footnote{\url{https://vizier.cds.unistra.fr/}} \citep{vizier2000} to search for all existing sources in the AllWISE catalogue \citep{wise2010,allwise2014}. A total of 110\,401  sources were retrieved including their IDs, positions and photometry in the $W1-W4$ and $JHK$ bands. Then, we applied the multicolour criteria scheme proposed by \citet{Koenig2014} to identify YSO candidates. This selection scheme is based on applying different cuts in colours and magnitudes in the WISE+2MASS bands to remove contaminants (Star-forming galaxies, Active Galactic Nuclei, and Asymptotic Giant Branch stars) and to select YSOs of Classes I, II, and Transition disks. The application of these criteria to our sample yielded a total of $1082$ YSOs, of which $139$ belong to Class I, $627$ to Class II, and $316$ are Transition disk sources. Table~\ref{tab_yso} (fully available online) presents a list of the selected YSOs, including their properties as derived in this work. An example colour-colour diagram is shown in Fig.~\ref{fig_yso_colorcolor}.
\begin{table*}
\caption{List of identified YSOs in the Dragonfish region. Here we show only a portion of the table for guidance regarding its form and content. The full version is available online and includes IDs from AllWISE, galactic and equatorial coordinates (J2000) in decimal degrees, magnitudes and errors in the WISE bands, object class according to \citet{Koenig2014}'s criteria, best model used for SED fitting, derived effective temperatures (K) and visual extinctions (mag), references for previously identified sources, and parallaxes and errors for the $135$ stars having counterparts in the Gaia DR3 catalogue.}
\label{tab_yso}
\centering
\small
\begin{tabular}{ccccccccccc}
\hline
ID(AllWISE) & RA & DEC & $W1$ & $W2$ & $W3$ & $W4$ & Class & Model & $T_{eff}$ & $A_V$ \\
\hline
J115840.15-631550.4 & $179.667315$ & $-63.264010$ & 12.433 & 12.225 & 9.906  & 7.442 & TrDisk \\
J115905.16-631608.3 & $179.771510$ & $-63.268990$   & 11.481 & 11.227 & 9.321  & 7.164 & Class~II \\
J115908.32-631559.7 & $179.784677$ & $-63.266603$ & 11.239 & 10.919 & 9.663  & 7.551 & Class~II \\
J115915.50-631310.4 & $179.814600$    & $-63.219573$ & 9.368  &  8.352 & 7.607  & 4.957 & Class~II & BT-Settl & 2600.0 & 2.0 \\
J115917.83-630950.1 & $179.824294$  & $-63.163930$ & 11.749 & 11.571 & 10.719 & 8.405 & TrDisk \\
\hline
\end{tabular}
\end{table*}
Of the selected sample, $323$ sources had already been reported either as YSOs or YSO candidates by other authors
\citep{Lumsden2013,Marton2016,Kuhn2021,Rimoldini2023,Zhang2023}.
Table~\ref{tab_yso} also includes the references for these previously identified YSOs. 
The remaining YSOs are new candidates, identified for the first time in this work.

\subsection{Physical properties of the selected YSOs}

\subsubsection{Distances}
\label{sec_distance}

\begin{figure}
\centering
\includegraphics[width=0.45\textwidth]{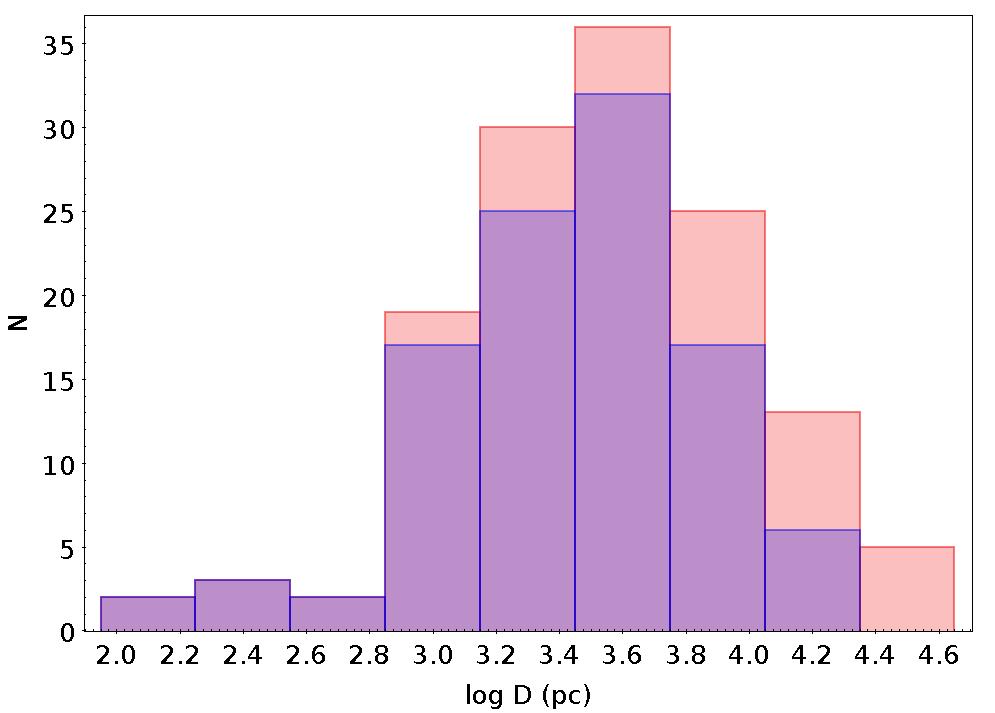}
\caption{Distribution of distances obtained from the parallaxes for the 135 stars that have counterparts in Gaia DR3. The purple histogram refers to sources having parallax errors smaller than the parallax values themselves.}
\label{fig_histo_distances}
\end{figure}

From the 1082 selected YSOs, there are 135 objects that have counterparts in the Gaia DR3 catalogue and therefore have available parallaxes. For these sources, we estimated their distances $D$ from their parallaxes, taking into account the global parallax offset of $-0.017$~mas reported by \citet{Lindegren2021}. In general, the obtained values of $D$ are distributed over a relatively large range of values, due in part to uncertainties in the parallaxes (see Fig.~\ref{fig_histo_distances}).
The most frequent value is found around $D \simeq 4000$~pc. If we consider parallaxes with relatively smaller errors (purple histogram in Fig.~\ref{fig_histo_distances}), the distribution changes slightly but the mode of the distribution always remains in the range $3000-5000$~pc. A distance of $D \sim 4 \pm 1$~kpc to the Dragonfish Complex is smaller than some previous estimations of $\sim 10-12$~kpc \citep{delaFuente2016,debuzier2022} but consistent with the $\sim 4-5$~kpc reported by other authors \citep{Moises2011,Rate2020}.

\subsubsection{Spectral energy distributions}

The SEDs of the 1082 selected YSOs were analysed using the Virtual Observatory SED Analyzer (VOSA) developed by the Spanish Virtual Observatory \citep[see details on the operation and limitations of VOSA in][]{Bayo2008}. VOSA is a tool that provides a friendly and flexible environment for finding the theoretical spectral model that best fits the observed photometric data. VOSA allows users to search and expand available photometry, choose from a list of models, and define parameter ranges to search for the best fit. For the fitting procedure, we first requested VOSA to expand the SEDs with all the photometry it could find. VOSA itself handles outlier rejection and, in case of finding different photometric values for the same filters, it calculates and uses an average value for the final SED. We then requested VOSA to fit the SEDs by minimising the reduced chi-square with the latest version of the BT-Settl models, which are based on the CIFIST photospheric solar abundances \citep{Caffau2011}. The effective temperature ($T_{eff}$) is left as a free parameter in the fitting process, which in the BT-Settl models ranges from $1200 \leq T_{eff} \leq 7000$~K. VOSA fits the points of the SED that have not been flagged with possible infrared excess, which it assumes to be around the $W1$ band. The fitting process in VOSA is not very sensitive to some parameters such as metallicity and $\log g$ \citep{Bayo2008}. We made several tests by fixing or constraining these parameters around the expected values and also leaving them completely free, and in general the resulting fits were not significantly affected. Eventually, we left $\log g$ as a totally free parameter and fixed the metallicity to the solar value, whereas the visual extinction ($A_V$) was allowed to vary in the range $0 \leq A_V \leq 10$~mag. For those sources for which the fitting process did not converge on reasonable solutions we also tried an independent fitting with the ATLAS9 Kurucz ODFNEW/NOVER models \citep{Castelli1997}. For these cases the temperature can vary in the range $3500 \leq T_{eff} \leq 50000$~K and the rest of the parameters are set to the same conditions employed for the BT-Settl fits. In any case, all 1082 SEDs and their fits were visually examined to verify the adequacy of the fits and solutions found by VOSA.

\begin{figure}
\centering
\includegraphics[width=0.45\textwidth]{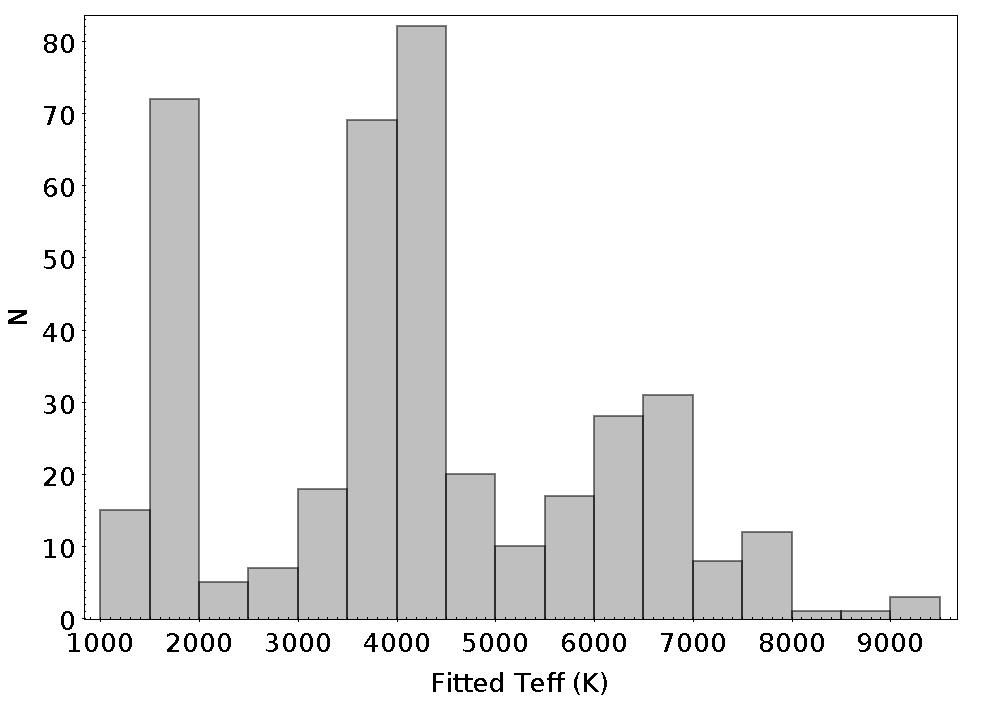}
\caption{Distribution of effective temperatures for the 399 sources whose SEDs could be fitted using VOSA (details in text).}
\label{fig_histo_teff}
\end{figure}

\begin{figure}
\centering
\includegraphics[width=0.5\textwidth]{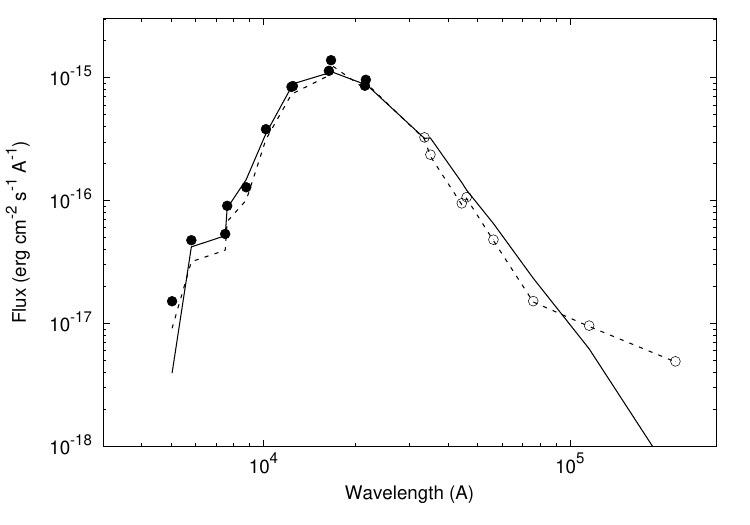}
\caption{Observed and best-fitted flux densities for one example source, J120022.63-631523.2, for which we obtained $T_{eff} = 1700$~K and $A_V = 0.5$, which is a Transition Disk object according to \citet{Koenig2014}'s criteria. Dashed line indicates the observed photometric data. Circles represent dereddened data, where solid circles denote data points that have been considered in the fitting process by VOSA. Solid lines indicate the best-fitted BT-Settl model. Some infrared excess is evident at wavelengths larger than $\sim 10~\mu$m.}
\label{fig_sed}
\end{figure}

For a total of 399 sources (37\%), the corresponding SED was well fitted with either BT-Settl models (89\% of the sources) or Kurucz models (11\%). The temperatures $T_{eff}$ and extinctions $A_V$ that provided the best fits are reported in Table~\ref{tab_yso}, and a histogram with the distribution of $T_{eff}$ values is shown in Fig.~\ref{fig_histo_teff}.
Around $\sim 50$\% of the sources have photospheric temperatures in the range $3000 \lesssim T_{eff} \lesssim 5000$~K, although there is also a significant population of cool stars (late Ms or brown dwarfs) with $T_{eff} \lesssim 2000$~K). An example SED for the first star belonging to this group in our Table~\ref{tab_yso} is shown in Fig.~\ref{fig_sed}, where we can see both the fit performed by VOSA using a BT-Settl model as well as the expected infrared excess likely produced by circumstellar dust.
We have not detected significant patterns or correlations of $T_{eff}$ or $A_V$ with the spatial distribution or with any other relevant physical variable. This population of cool stars is subject to ongoing investigation.

\subsection{Spatial distribution}

\begin{figure*}
\centering
\includegraphics[width=\textwidth]{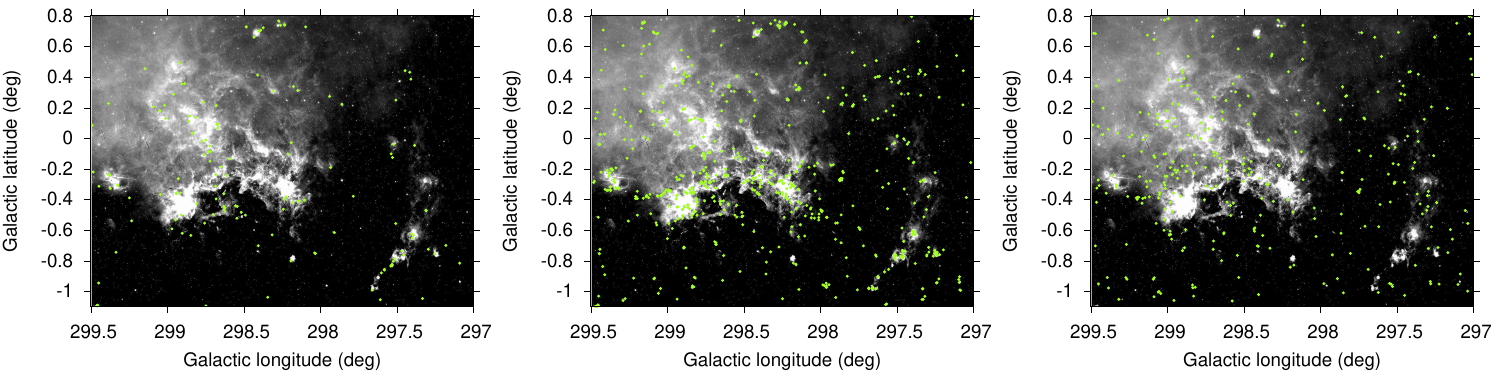}
\caption{Spatial distribution of YSO candidates selected according to \citet{Koenig2014}'s criteria overlaid on the 8~$\mu$m band map from Spitzer IRAC. Green dots are sources that fulfil criteria of Class I (left panel), Class II (center panel) or Transition disk (right panel) objects.}
\label{fig_yso_distribucion}
\end{figure*}

The spatial distribution of the selected YSOs, overlaid on the Spitzer map at $8~\mu$m, is shown in Fig.~\ref{fig_yso_distribucion}.
At first glance, younger classes (I and II) seem to follow the distribution of gas and dust exhibiting some level of clumpiness, whereas Transition disk sources tend to be more homogeneously spread through the region. In order to objectively quantify the clumpiness we use the so-called correlation dimension ($D_c$), which is suitable for analysing distributions of point sources. The correlation dimension measures the variation (as $r$ increases) of the probability that two randomly chosen points are separated by a distance smaller than $r$ \citep{Grassberger1983}. For homogeneous point distributions in space it is expected that $D_c=3$, whereas in a plane $D_c=2$. If the points are distributed following fractal patterns then $D_c<3$ in the space or $D_c<2$ in the plane. Here, we use a previously developed and calibrated algorithm that estimates $D_c$ in a precise and accurate manner \citep{Sanchez2007b,Sanchez2008}.
The algorithm constructs the minimum-area convex polygon to delimit the sample and avoid common problems at large scales (whole sample scale). On the other hand, at spatial scales of the order of the mean distance to the nearest neighbour, the distribution looks like a set of isolated points and the obtained $D_c$ values tend to zero \citep{Smith1988}. Our algorithm uses suitable criteria to eliminate poorly estimated data (i.e., bad sampling) and thus to avoid these small-scale issues. 
Additionally, it applies bootstrapping techniques to estimate an uncertainty associated to $D_c$. The results from applying this algorithm are presented in Table~\ref{tab_dc}. 
\begin{table}
\caption{Calculated correlation dimension ($D_c$) and corresponding three-dimensional fractal dimension ($D_f$) for the different classes of YSOs shown in Fig.~\ref{fig_yso_distribucion}. The number of data points ($N$) for each sample is also shown.}
\label{tab_dc}
\centering
\small
\begin{tabular}{c c c c}
\hline
YSO Class & $N$ & $D_c$ & $D_f$ \\
\hline
All classes & $1082$ & $1.74 \pm 0.02$ & $1.9-2.0$ \\ 
Class I & $139$ & $1.55 \pm 0.07$ & $1.6-1.7$ \\
Class II & $627$ & $1.63 \pm 0.03$ & $1.7-1.8$ \\
Transition disk & $316$ &  $1.84 \pm 0.04$ & $2.1-2.3$ \\
\hline
\end{tabular}
\end{table}
The estimation of the corresponding three-dimensional fractal dimension $D_f$ is made based on the simulations and results in \citet{Sanchez2008}. 

The obtained $D_f$ values reveal a certain evolutionary process. Classes I and II exhibit approximately the same value of $D_f \simeq 1.7 \pm 0.1$. The slightly smaller value of $D_c$ (and $D_f$) for the distribution of Class I objects is likely related to the relatively small number of sources ($N=139$), because it has been shown that below $N \sim 200$ 
the retrieved value of $D_c$ tends to be smaller than the actual fractal dimension \citep[see Fig. 2 in][]{Sanchez2008}. In contrast, the more evolved Transition Disk objects show a significantly larger dimension with $D_f \simeq 2.2 \pm 0.1$. There is evidence for such evolutionary effect, where the initial hierarchical and clumpy structure gradually disappears over time \citep{Elmegreen2018}. In external galaxies, the clumpy structure in the distribution of star formation sites has been observed to change towards smoother distributions as ages increase \citep{Gieles2008,Sanchez2008,Bastian2009,Bonatto2010,Sanchez2010,Menon2021}. At these scales ($\gtrsim 10^3$~pc), the underlying cause may lie in non-turbulent motions acting at a galactic level, on scales on the order of or larger than the scale height of galactic disks \citep[see discussions in][]{Sanchez2010,Menon2021}. In the case of star clusters, the observed initial substructures also seem to dissipate with age \citep{Schmeja2008,Sanchez2009,Ballone2020,Daffern2020}, but on those spatial scales ($\sim 10$~pc) other physical processes may play important roles and the initial fractal structure is expected to be lost rapidly \citep{Elmegreen2018}, either diluting into more homogeneous distributions in gravitationally unbound clusters or concentrating into radial density distributions in bound clusters. Nevertheless, the associated time scale is not clear and some works suggest that cluster disruption may be a very slow process ($\gtrsim 10$~Myr) in some cases \citep{Hetem2019}. In any case, different young clusters may reflect the initial structure of the different clouds from which they formed, and these conditions do not necessarily have to be the same. Therefore, reported correlations between clumpiness and age for large samples of cluster could be contaminated by differences in initial conditions and not correspond to any evolutionary effect. In the case of the Dragonfish complex, we focus on spatial scales of the order of $\sim 100$~pc, where we detect a significant difference between the distributions of younger and more evolved stars, marking an evolutionary effect. The underlying physical process is still not clear but it may be related to the random stellar motion effect discussed by \citet{Elmegreen2018}, in which the initial clumpy structure disappears as stars age due to random turbulent velocity fields acquired at birth.

\subsection{Hierarchy of structures}

The previous analysis points out that YSOs are distributed in a clumpy manner showing some degree of substructure at different spatial scales. In this section, we address this issue by searching for density substructures in the YSO sample. For this, we
apply the algorithm OPTICS \citep[Ordering Points To Identify the Clustering Structure;][]{Ankerstetal99} to perform a global analysis of the density structure of YSOs and retrieve a hierarchy of subclusters. OPTICS extends the clustering algorithm DBSCAN \citep{Esteretal96} to provide a global analysis of the density structure within a region and is especially suited for samples with large density variations. DBSCAN 
groups points into clusters based on
a density associated with two parameters: a minimum number of points $N_{min}$ and a spatial scale $\varepsilon$. For each point, the scale $\varepsilon$ defines a neighbourhood, and $N_{min}$ sets a density requirement for the neighbourhood. DBSCAN identifies clusters as composed of two kinds of points: core points satisfy the density requirement, while border points belong to the $\varepsilon$-neighbourhood of a core point but do not satisfy the density requirement themselves. The rest of the points are labelled as noise. In OPTICS, only the $N_{min}$ parameter is fixed and the concept of reachability distance is introduced. We can intuitively interpret the reachability distance between a core point and another point as the minimum distance needed for the second point to be in the $\varepsilon$ neighbourhood of the core point, fulfilling the density threshold. OPTICS is not strictly a clustering algorithm, but an analytical tool whose main output is the reachability plot. The reachability plot shows the reachability distance of a reordered sample of points in a diagram where consecutive points are close and the clusters appear as dents or valleys. Based on the reachability plot, we can extract clusters either considering a height $\varepsilon$ threshold (obtaining a clustering equivalent to DBSCAN with that $\varepsilon$) or with a slope $\xi$ threshold (where clusters have a specific density ratio to their surroundings). In this work, we use the second approach, as it can detect a hierarchy of structures nested within each other. We set a high value $N_{min}=20$ to limit the noise in the reachability plot and tested values of $\xi \in [0.01,~ 0.1]$, finally choosing $\xi=0.035$ as a good compromise between the detection and the reliability of the clusters retrieved.

\begin{figure}
\centering
\includegraphics[width=0.5\textwidth]{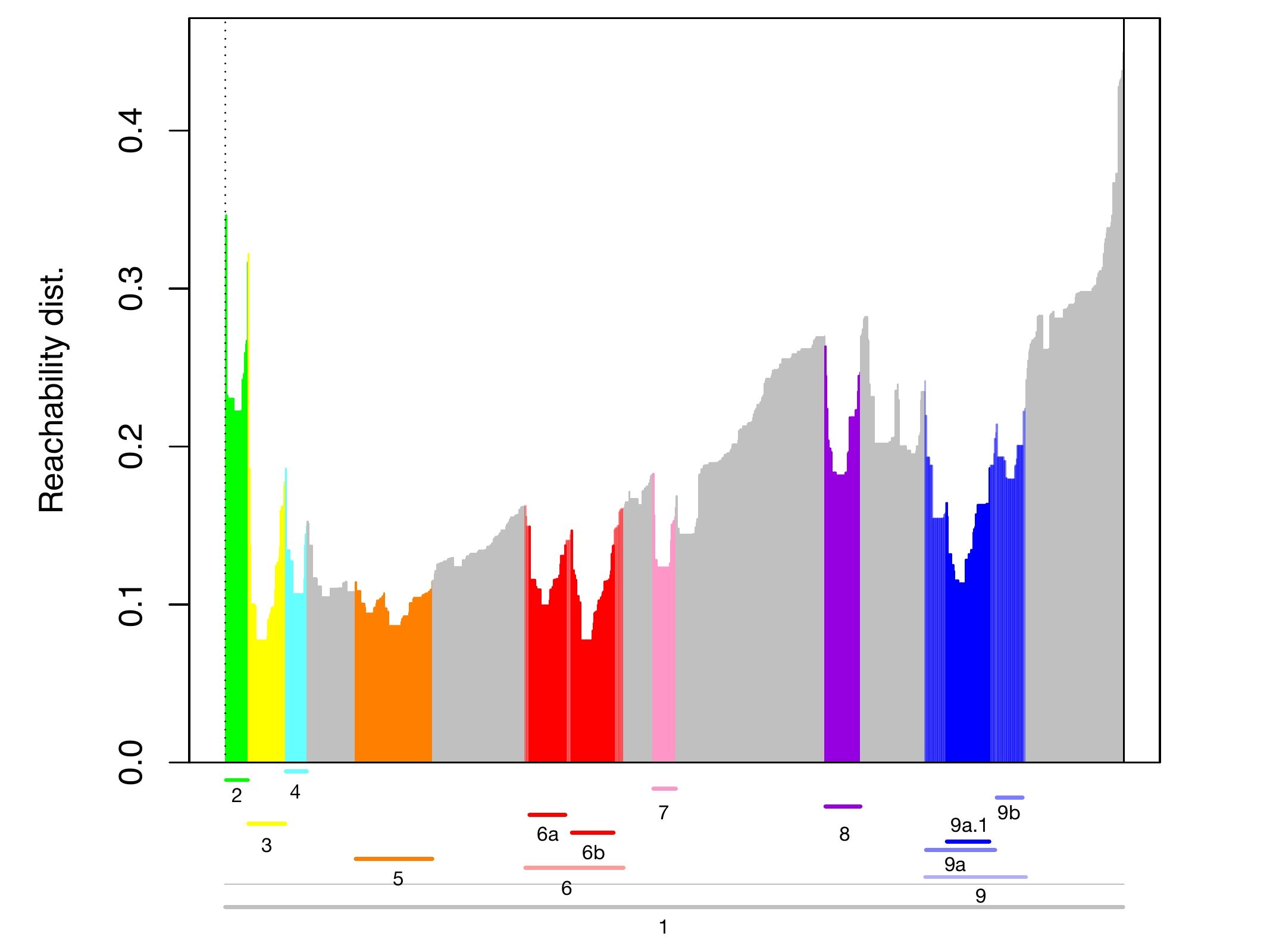}
\caption{Reachability plot obtained by OPTICS. The retrieved structures are colour-coded and labelled on the X-axis.}
\label{fig_reachplot}
\end{figure}

\begin{figure*}
\centering
\includegraphics[width=1.035\textwidth]{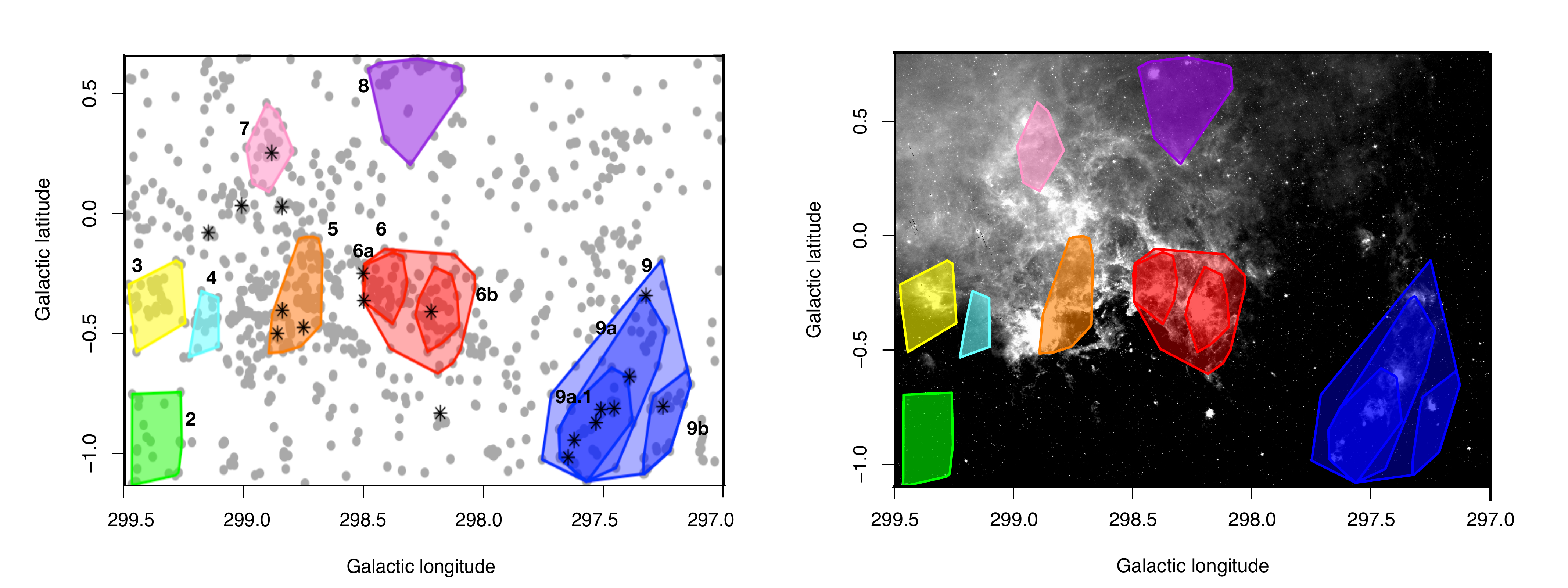}
\caption{Left panel: YSOs (grey dots) and convex hulls of the retrieved structures, labelled and coloured as in Fig~\ref{fig_reachplot}. Black asterisks are candidate or confirmed star clusters taken from Table~7 in \citet{delaFuente2016}. Right panel: Overlap of the convex hulls of structures and the 8~$\mu$m emission map from Spitzer IRAC.}
\label{fig_clusters}
\end{figure*}

Figures~\ref{fig_reachplot} and \ref{fig_clusters} show the reachability plot and a map with the convex hulls of the structures retrieved by OPTICS, in both cases labelled and colour-coded.
In Table~\ref{tabla_clusters} we display the obtained characteristics of each structure.
\begin{table*}[ht]
\caption{List of the retrieved clustered structures and substructures indicating their IDs, number of sources ($N$) and the assigned level for each structure, the corresponding mean right ascension and declination in degrees, the equivalent radius in degrees (the radius of a circle having the same area of the associated convex hull), the number of sources for each YSO class, the fraction $f_{CI}$ of Class~I sources with respect to the total of YSOs, and cluster candidates in the region listed by \citet{delaFuente2016}.}
\label{tabla_clusters}
\centering
\small
\begin{tabular}{ccccccccccl}
\hline
ID  & N & Level & Mean RA & Mean DE & Radius & Class~I & Class~II & TrDisk & $f_{CI}$ & Candidates\\ 
\hline
2 &  27 &   0 & 184.90 & -63.62 & 0.24 &   6 &  13 &   8 & 0.22 &\\ 
\hline
3 &  45 &   0 & 185.02 & -62.94 & 0.21 &   8 &  27 &  10 & 0.18 & \\ 
\hline
4 &  26 &   0 & 184.54 & -63.01 & 0.13 &   1 &  13 &  12 & 0.04& \\ 
\hline
5 &  93 &   0 & 183.71 & -62.86 & 0.23 &  11 &  71 &  11 & 0.12 & Mercer 30, 31\\ 
  &  &  &  &  &  &  &  &   &  & DBSB129\\ 
\hline
6 & 119 &   1 & 182.67 & -62.80 & 0.36 &   8 &  90 &  21 & 0.07 & \\ 
6a &  44 &   0 & 182.94 & -62.77 & 0.17 &   0 &  35 &   9 & 0.00 & La Serena 24\\
   &  &   &  &  & &   &   &  & & VVV Cl011\\ 
6b &  51 &   0 & 182.46 & -62.83 & 0.18 &   7 &  41 &   3 & 0.14 & DBSB74\\ 
\hline
7 &  28 &   0 & 184.19 & -62.21 & 0.18 &   4 &  14 &  10 & 0.14 & La Serena 28\\ 
\hline
8 &  43 &   0 & 182.98 & -61.83 & 0.28 &  12 &  25 &   6 & 0.28 &\\ 
\hline
9 & 122 &   2 & 180.61 & -63.08 & 0.50 &  22 &  74 &  26 & 0.18 &\\ 
9a &  84 &   1 & 180.74 & -63.08 & 0.36 &  18 &  51 &  15 & 0.21 & La Serena 18\\ 
   &  &   &  &  &  & &   &  & & Mercer 28\\ 
9a1 &  52 &   0 & 180.85 & -63.21 & 0.25 &  14 &  33 &   5 & 0.27 & La Serena 19, 20, 22\\ 
    &   &    &  &  &  & & &   &  & Mercer 29\\ 
    & &   & &  &  &   & && & New Candidate\\ 
9b &  32 &   0 & 180.23 & -63.14 & 0.18 &   4 &  20 &   8 & 0.12 & La Serena 17\\ 
\hline
\end{tabular}
\end{table*}
The whole sample is itself
detected as a single structure by OPTICS,
tagged and displayed in grey as structure 1 in Fig.~\ref{fig_reachplot} but not shown in Fig.~\ref{fig_clusters} and Table~\ref{tabla_clusters} for the sake of clarity.
The YSO sample exhibits a rich hierarchical structure, as expected from the previous fractal analysis. There are 8 main structures, and amongst them, structures 6 and 9 contain nested substructures. We assign to each structure a level, defined recursively by the number of substructures it contains: structures of level 0 do not contain any other, structures of level 1 contain structures of level 0, structures of level 2 contain structures of level 1, and so on. We find 10 structures of level 0 (namely 2, 3, 4, 5, 6a, 6b, 7, 8, 9a1, and 9b), 2 of level 1 (6 and 9a), 1 of level 2 (9) and 1 of level 3 (structure 1,
i.e. the whole sample).
By considering a distance of $\sim 4$~kpc (see Section~\ref{sec_distance}) and the average equivalent sizes of the structures for each level (shown in Table~\ref{tabla_clusters}), we obtain that 
the spatial scales of the structures are of the order of 15, 25, and 25~pc, typical of star-forming regions.

Table~\ref{tabla_clusters} also shows the correspondence between our detected structures and star cluster candidates listed in Table~7 of \citet{delaFuente2016}. Out of their 19 candidates, 15 are found within our structures.
Our structure 3 includes the location of the H~II region RCW~64 \citep{Caswell1987} that \citet{delaFuente2016} consider as a foreground region. However, there are 4 objects inside structure 3 with reliable parallaxes from Gaia DR3 which place this structure at a distance of $\sim 3.8 \pm 0.6$~kpc, consistent with our estimation for the whole region (see \ref{sec_distance}).
Structures 5,6, and 9 contain three or more known star cluster candidates, and, in the case of the most complex structure 9, its higher level substructure 9a1 contains 5 candidates and displays a clear elongated shape. Even though environmental factors must play some role, the fraction of less evolved Class~I sources to the total number of YSOs may yield some hints on the age of the structure \citep{Evans2009}.
In fact, the variations in the spatial distribution of objects of different evolutionary stages have been used to investigate the star formation history within specific regions
\citep[see e.g.][]{Sung2009,Venuti2018,Nony2021,Flaccomio2023}
For the whole sample, the ratio Class~I/YSO is $12.9\%$ which is broadly consistent with the value $14.4\%$ obtained considering only the clustered structures. However, we also identified structures with ratios both larger than $20\%$ and smaller than $5\%$. Structures 2, 8 and 9a (in particular 9a1) have significantly large Class~I/YSO ratios that suggest a higher level of recent star formation, whereas structures 6a and 4 show significantly low ratios, pointing to areas where the star formation activity has declined. These results suggests a complex and varied star formation history in the Dragonfish complex, comprising different events spanning several Myr.

\section{Comparison of gas and YSO distributions}
\label{sec_discusion}

A primary goal of this work is to compare the distribution of gas and dust in the Dragonfish region with the distribution of young stars that were born from this gas. Given that we are using the same characterisation tool (the fractal dimension) for gas and for stars, in principle we expect to obtain nearly the same $D_f$ value for both components. The reason is that, at least in the case of ideal monofractal clouds with dimension $D_f$, the high-density peaks where star formation preferentially takes place are distributed following patterns with the same underlying dimension $D_f$. On the contrary, our results clearly indicate a scenario in which the distribution of younger objects is much more clumpy ($D_f \simeq 1.7$) than the material from which they are forming ($D_f \simeq 2.6-2.7$).

Evidence supporting or contradicting either of these scenarios is far from being conclusive. There is a lack of works that compare the clustering strength of gas and new-born stars in a direct, quantitative and self-consistent way. At spatial scales smaller than $500$~pc, the galaxy M33 is on average more fragmented and irregular than the Milky Way, but its bright young stars are distributed following nearly the same fractal patterns as the molecular gas, both having $D_f \lesssim 1.9$ \citep{Sanchez2010}. In contrast, for the galaxy NGC~7793, \citet{Grasha2018} found that on average star clusters are distributed with a stronger clustering degree than giant molecular clouds over the range $40-800$~pc; nevertheless, they also found approximately the same degree of clustering when comparing the most massive molecular clouds with the youngest and most massive star clusters. In any case, at spatial scales of the order of the disk scale height, the structure of the interstellar medium generated by turbulent motions may be somehow affected by large-scale galactic dynamics, as discussed in \citet{Sanchez2010} and \citet{Menon2021}. \citet{Gregorio2015} calculated the clustering in a sample of young star clusters using the $Q$-parameter \citep{Cartwright2004} and also estimated the perimeter-area-based dimension $D_{per}$ from visual extinction maps in the direction of such clusters. In general, they found that substructures observed for the clusters were very similar to the fractal characteristics of the clouds, although an accurate comparison of the three-dimensional fractal dimension $D_f$ could not be made because projection effects were not properly taken into account. \citet{Parker2015} performed a direct comparison of the spatial distributions of stars and gas in numerical simulations of molecular clouds using the $Q$-parameter for both stars and gas. Interestingly, they found that formed stars follow a distribution highly substructured with a value of $Q\sim 0.4-0.7$, which corresponds to approximately $D_f \sim 1.8-2.3$ \citep[see Fig.~7 in][]{Sanchez2009}, whereas the gas from which stars form had $Q\sim 0.9$, indicating a smooth, concentrated distribution of matter. These results should nevertheless be treated with some caution because, as indicated by \citet{Parker2015}, the $Q$-parameter may not be an optimal tool for measuring the spatial distribution of gas, since the pixelated image must be previously converted into a point distribution.

Here, we have found direct and self-consistent evidence that the clustering degree of newly born stars in the Dragonfish region is significantly higher than that of the parent cloud from which stars are forming. 
We have mentioned that resolution issues (relatively big pixel sizes) could make gas maps look smoother than they actually are, an effect that would not occur for the distribution of stars. However, this is likely not the cause of the observed difference between stars and gas because resolution effects and other factors (signal-to-noise ratio, cloud opacity) have already been calibrated and accounted for to accurately infer $D_f$ from $D_{per}$ \citep{Sanchez2005,Sanchez2007a}. If this discrepancy is real, then there could be two possible explanations.
On the one hand, the denser gas that is forming stars could be actually clumpier than the distribution of gas throughout the entire region, in concordance with a possible multifractal scenario that has been proposed for the interstellar medium \citep{Chappell2001}. The approach used in this work aims to avoid this issue by focusing on spatial scales of the same order for both the cloud structure and the YSO distribution. On the other hand, the degree of clumpiness may somehow increase during the star formation process. Although some simulations \citep[e.g.,][]{Parker2015} seem to support this possibility, the physical mechanisms driving this behaviour remain unclear. In simulations of fractal star clusters, \citet{Goodwin2004} demonstrated that an initially homogeneous cluster can develop substructures if it is born with some coherence in the initial velocity field. It is not clear, however, whether such kind of processes could operate at the spatial scale of a whole star-forming complex.

The relationship between the spatial distributions of gas and formed stars remains as an intriguing, and so far unresolved, issue. The problem is non-trivial because the conversion from gas to stars, i.e. the star formation process, involves many physical mechanisms interacting at different spatial scales. Moreover, the formation of stars does not occur synchronously throughout the entire cloud, and the first-formed stars can interact with the surrounding gas modifying its properties and also affecting the distribution of subsequent stars.

\section{Conclusions}
\label{sec_conclusion}

In this paper, we present a systematic and detailed study of the Dragonfish star-forming region. On the one hand, we used different emission maps to characterise the distribution of gas and dust using fractal analysis. The three-dimensional fractal dimension obtained for the Dragonfish Nebula was $D_f \simeq 2.6-2.7$, a value that agrees very well with previously reported fractal dimensions for other molecular clouds, namely Orion, Ophiuchus and Perseus. On the other hand, we used photometric information from the AllWISE catalogue to select and study a total of 1082 YSOs in this region. From the parallaxes measured by the Gaia mission for 135 of these sources we derived a distance of $D \sim 4 \pm 1$~kpc to the Dragonfish Complex and from the SED fitting to theoretical models we also determined photospheric temperatures and visual extinctions for 399 sources. Regarding the spatial distribution of YSOs, we identified a clumpy a hierarchical assembly of structures and substructures and, moreover, we found that the clumpy structure of younger Class~I and Class~II sources tends to disappears for the more evolved sources (Transition Disks), suggesting some kind of evolutionary effect. Interestingly, our fractal analysis clearly show that the distribution of younger objects is much more clumpy ($D_f \simeq 1.7$) than the distribution of gas from which they formed ($D_f \simeq 2.6-2.7$). Although some simulations \citep[e.g.,][]{Parker2015} seem to support the possibility that newly formed stars exhibit a more clumpy structure than that of their parent cloud, the physical mechanisms behind this behaviour remain unclear. In order to clarify this issue, it would be helpful to use strategies such as the one proposed in this work, in which suitable and well-calibrated tools are used to simultaneously quantify the structure of both gas and stars in a relatively large sample of star-forming complexes.

\begin{acknowledgements}
We want to thank the referee for his/her helpful comments, which improved this paper.
We acknowledge financial support from Universidad Internacional de Valencia (VIU) through project VIU24003. The work of EN was supported by project VIU24007, funded by the reasearch center ESENCIA of VIU. JBC was supported by projects PID2020-117404GB-C22, funded by MCIN/AEI, CIPROM/2022/64, funded by the Generalitat Valenciana, and by the Astrophysics and High Energy Physics programme by MCIN, with funding from European Union NextGenerationEU (PRTR-C17.I1) and the Generalitat Valenciana through grant ASFAE/2022/018.
This research made use of Montage, that it is funded by the National Science Foundation under Grant Number ACI--1440620, and was previously funded by the National Aeronautics and Space Administration's Earth Science Technology Office, Computation Technologies Project, under Cooperative Agreement Number NCC5--626 between NASA and the California Institute of Technology. We have made extensive use of VOSA, developed under the Spanish Virtual Observatory project funded by MCIN/AEI/10.13039/501100011033/ through grant PID2020-112949GB-I00. We also have used the tool TOPCAT \citep{Taylor2005} and the NASA’s Astrophysics Data System.
\end{acknowledgements}

\bibliographystyle{aa}
\bibliography{Dragonfish}

\end{document}